\documentclass{sig-alternate}

\usepackage{graphicx}
\usepackage{balance}
\usepackage{times}

\newcommand{\QED}{\mbox{\rule[0pt]{1.0ex}{1.0ex}}}
\def\boxend{\hspace*{\fill} $\QED$}

\newtheorem{example}{Example}
\newtheorem{theorem}{Theorem}
\newtheorem{problem}{Problem}
\newtheorem{lemma}{Lemma}

\newcommand{\nop}[1]{}

\newcounter{line}
\begin{document}


\title{Fast and Quality-Guaranteed Data Streaming in Resource-Constrained Sensor Networks}

\numberofauthors{3}

\author{
 \alignauthor
   Emad Soroush \\
   \affaddr{Dept. of Computer Science} \\
   \affaddr{University of Victoria} \\
   \affaddr{Victoria, BC Canada V8W 3P6} \\
   \email{soroush@csc.uvic.ca}
 \alignauthor
   Kui Wu \\
   \affaddr{Dept. of Computer Science} \\
   \affaddr{University of Victoria} \\
   \affaddr{Victoria, BC Canada V8W 3P6} \\
   \email{wkui@csc.uvic.ca}
 \alignauthor
   Jian Pei \\
   \affaddr{School of Computing Science} \\
   \affaddr{Simon Fraser University} \\
   \affaddr{Burnaby, BC Canada V5A 1S6} \\
   \email{jpei@cs.sfu.ca}
}

\maketitle

\thispagestyle{empty}

\section*{ABSTRACT}
In many emerging applications, data streams are monitored in a
network environment. Due to limited communication bandwidth and
other resource constraints, a critical and practical demand is to
online compress data streams continuously with quality guarantee.
Although many data compression and digital signal processing methods
have been developed to reduce data volume, their super-linear time
and more-than-constant space complexity prevents them from being
applied directly on data streams, particularly over
resource-constrained sensor networks. In this paper, we tackle the
problem of online quality guaranteed compression of data streams
using fast linear approximation (i.e., using line segments to
approximate a time series). Technically, we address two versions of
the problem which explore quality guarantees in different forms. We
develop online algorithms with linear time complexity and constant
cost in space. Our algorithms are optimal in the sense they generate
the minimum number of segments that approximate a time series with
the required quality guarantee. To meet the resource constraints in
sensor networks, we also develop a fast algorithm which creates
connecting segments with very simple computation. The low cost
nature of our methods leads to a unique edge on the applications of
massive and fast streaming environment, low bandwidth networks, and
heavily constrained nodes in computational power.\nop{ (e.g., tiny sensor
nodes).} We implement and evaluate our methods in the application of
an acoustic wireless sensor network.

\section*{Categories and Subject Descriptors}
C.3 [\textbf{Computer Systems Organization}]: Special-Purpose and  Ap\linebreak plication-Based Systems; G.1.2 [\textbf{Mathematics of Computing-\linebreak Numerical  Analysis}]: Approximation-\textit{Linear approximation}
\section*{General Terms} Algorithms, Design, Performance
\section*{Keywords} Wireless Sensor Networks, Data Streaming, Linear Approximation

\section{Introduction}\label{sec:intro}

In many emerging applications, massive data streams are monitored in
a network environment. For example, large sensor networks are
extensively used in wildlife monitoring, road traffic monitoring,
and environment surveillance. Each sensor generates a data stream
where new data entries (i.e., new readings) keep arriving in a
continuous manner. In order to aggregate and analyze the massive
streaming data under monitoring, it is often required to transmit
the data streams in the network. Due to often limited communication
bandwidth and other resource constraints, online compressing data
streams continuously with quality guarantee rises as a natural,
critical and practical demand in those applications.

\begin{example}[Motivation]\label{ex:motivation}\em
We, the authors of this paper, are building an acoustic monitoring
system using wireless sensor networks. Sensor nodes are deployed in
a target area, while each node contains an acoustic sensor which
samples sound signals continuously. The sensor nodes are connected
by a wireless network.

The acoustic monitoring system has many applications. An appealing
scenario is towards ``smart conference hall.'' By analyzing the data
collected from an acoustic monitoring system deployed in a large
conference place, we can identify and locate speakers as well as
some of their activities. The information can be used to adjust the
equipment such as the light system, the microphone system, the video
monitoring system, and the air conditioning system. Another
potential application is bird surveillance in wildness. By analyzing
the bird sound collected using such a sensor network, ornithologists
can study the distribution of birds and their behavior patterns.

Wireless sensor nodes which integrate sensors, processors, memory
and wireless transceivers often are small and have only very limited
computational power and communication bandwidth. For instance, the
Chipcon radio chip in the broadly-used MICA2 motes ~\cite{MICA2} has
the maximum transmission power of $27$ mA and the maximum bandwidth
of $38$~kbps.

In our acoustic monitoring system, we use MICA2 motes. One technical
challenge is that, although a sensor can sample the acoustic signals
frequently, the acoustic data stream cannot be sent out in time due
to the low bandwidth radio channel. Specifically, in order to make
the data analysis useful, we need to sample human voice with the
normal sampling rate of $8$ kHZ and 16 bits per sample. This
sampling mode requires the bandwidth of $128$ kbps for 1 channel
(mono) voice, which greatly exceeds the maximum bandwidth of $38$
kbps that an MICA2 mote can support.
In addition, we cannot temporarily store a
large number of samples since the memory size of MICA2 motes is only
$512$ kb. The only technical solution to the bottleneck is to online
compress data streams continuously and send out the compressed
streams instead of the original streams through the network. Sending
compressed streams can also reduce the power consumption of sensors
on communication, and thus extend lifetime of sensors. In large
environmental surveillance sensor networks, recharging or replacing
batteries of sensor nodes is often very difficult or even impossible
after the sensors are deployed. \boxend
\end{example}

Many data compression and digital signal processing methods have
been developed to reduce data volume, such as Fourier
transform~\cite{Fourier1}, discrete cosine
transform~\cite{DCT1}, Wavelets~\cite{chan99efficient}, linear predictive coding (LPC)~\cite{LPC}, etc.
However, those methods cannot be applied to data stream compression
in sensor networks due to the high cost of those methods in time and
space. Moreover, sensor nodes like MICA2 motes only have very
limited computational power. For example, only simple arithmetic
operations are supported by TinyOS~\cite{Cull}, the operating system
for MICA2 motes. Although it is possible to implement a mathematical
module to calculate essential functions like sinusoid and
exponential functions or use dedicated DSP chips for audio
processing and compression, such complex modules are highly
undesirable due to the limited memory size and computational
capacity of MICA2 motes as well as the extra energy cost of
dedicated DSP chips.

In this paper, we tackle the problem of online compression of data
streams in the application context of sensor networks. Particularly,
we aim at the fast linear approximation methods (i.e., using line
segments to approximate a time series) with quality guarantee. We
make the following contributions.

First, we model the piecewise linear approximation problem properly
for data streams. Different from the conventional situations where
the whole time series to be compressed and the required compression
rate can be specified, a data stream is potentially unlimited, and
the distribution is often unpredictable. We propose the
error-bounded piecewise linear approximation problem to tackle those
challenges. Second, we present fast online solutions with linear
time complexity and constant cost in space. Our algorithms are
optimal in the number of segments used to approximate a (potentially
unlimited) time series. In other words, our algorithms create the
minimum number of line segments {\em even without knowing the future
incoming data}.\nop{Moreover, our algorithms have an approximation
factor of $2$ to the maximum compression ratio using piecewise
linear approximation.} To the best of our knowledge, we are the
first to successfully devise algorithms with such strong guarantees.
Third, to address the computational challenges in sensor nodes, we
develop another online approximation algorithm that is particularly
tailored for tiny sensor devices by requiring only very simple
computation. The low cost nature of our methods leads to a unique
edge on the applications of massive and fast streaming environment,
low bandwidth networks, and heavily constrained nodes in
computational power (e.g., tiny sensor nodes). Last, we implement
and evaluate our methods in the application of an acoustic wireless
sensor network. Our empirical evaluation clearly shows that our
methods are highly feasible for resource-constrained wireless sensor
networks.

The rest of the paper is organized as follows. In
Section~\ref{sec:prob}, we formulate and analyze the problem, and
review the related work. Two online algorithms are developed in
Section~\ref{sec:online}, and their optimality is studied in
Section~\ref{sec:optimal}. In Section~\ref{sec:PLAZA}, we design an
online approximation algorithm which is more economic in computation
for tiny sensors. We report our implementation and evaluation of the
proposed methods in an acoustic wireless sensor network in
Section~\ref{sec:exp}. The paper is concluded in
Section~\ref{sec:con}.

\section{Problem Definition and Related Work}\label{sec:prob}

In this section, we propose the error-bounded piecewise linear
approximation problem for data streams. We also review the related
work.

\subsection{Problem Formulation}

Piecewise linear approximation (PLA) is an effective method to
compress a time series. A numeric data stream can be treated as a
potentially unlimited time series. Thus, it is natural to explore
whether we can compress a numeric data stream using the piecewise
linear approximation method.

\begin{figure}[t]
  \centerline{\includegraphics[width=60mm]{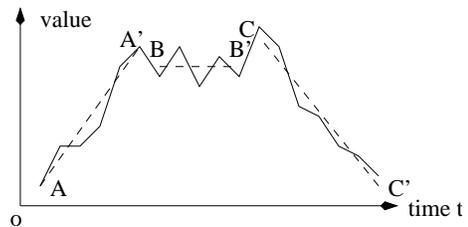}}
  \caption{\label{fig:pla} Piecewise linear approximation.}
\end{figure}

Let $X=x_1 \cdots x_n$ be a time series of $n$ points, and $x_i$ $(1
\leq i \leq n)$ be the value of the $i$-th point of $X$. A {\em
(line) segment} is a tuple $s=((i, y_i), (j, y_j))$ where $i < j$
and $(i, y_i)$ and $(j, y_j)$ are two endpoints. $[i, j]$ is called
the {\em range} of $s$.

Given a time series $X$, PLA uses a set of line segments as the
approximation of the time series. Figure~\ref{fig:pla}
elaborates the general idea, where three line segments, $AA'$,
$BB'$, and $CC'$, are used to approximate a time series. A line
segment $s=((i, y_i), (j, y_j))$ approximates the $k$-th point $(i
\leq k \leq j)$ of the time series by value
\[\tilde{x}_k=y_i+\frac{k-i}{j-i}(y_j - y_i).\] The compression comes
from that the number of line segments used for approximation can be
much smaller than the number of points in the time series. In the
figure, the time series has $18$ points. three segments are used to
approximate the time series, and each segments has $2$ endpoints.
Thus, the 3 line segments only need $6$ points to represent. A
compression ratio of $3$ is achieved. Generally, the endpoints in
the segments are not necessarily positioned at some points in the
time series (e.g., $B$, $B'$, and $C'$ in the figure).

Formally, a set of segments $\tilde{X}=\{s_1, \ldots, s_m\}$ is a
piecewise linear approximation of $X$ if (1) $s_1, \ldots,
s_m$ are segments; and (2) for each index $i$ $(1 \leq i \leq n)$,
$i$ is either in the range of exactly one segment in $\tilde{X}$, or
there exist two segments $s, s' \in \tilde{X}$ such that $s$ and
$s'$ share the same endpoint at index $i$. Clearly, using the
segments, for every index $i$, $\tilde{X}$ can give a value
$\tilde{x}_i$ to approximate $x_i$.

PLA for static time series has been well studied
(e.g.,~\cite{Dunh86, Good94, keogh98enhanced, qu98supporting}). Most of the previous studies
address an optimization problem as follows.

\begin{problem}[Conventional PLA problem]
Given a time series $X$ of $n$ points and a number $m < n$, find a
set of $m$ segments as a piecewise linear approximation of $X$ such
that the approximation error is minimized. \boxend
\end{problem}

Unfortunately, solutions to the conventional PLA problem are not
applicable to data streams. A data stream is potentially unlimited.
It is impossible to know in advance the number of points in the
stream or to specify the number of segments to be used for
approximation. To tackle the stream compression problem, in this
paper, we turn to the {\em error-bounded PLA problem}.

\begin{problem}[Error-bounded PLA problem]
Given an error measurement function $err()$ such that $err(X,
\tilde{X})$ gives the error that a PLA $\tilde{X}$ approximates $X$.
Let $\epsilon$ be a user-specified error bound. $\tilde{X}$ is
called an $\epsilon$-PLA of $X$ if $err(X, \tilde{X}) \leq
\epsilon$. An $\epsilon$-PLA $\tilde{X}$ of $X$ is {\em optimal} if
$|\tilde{X}|$ (i.e., the number of segments in $\tilde{X}$) is
minimized. \boxend
\end{problem}

We propose two error measurement functions meaningful for data
streams.

First, the {\em max-err} function captures the maximal error between
$X$ and $\tilde{X}$ at any index. That is,
\[maxerr(X, \tilde{X})=\max_{i=1}^n \{|x_i-\tilde{x}_i|\}\] With
potentially unlimited streams, using the max-err function, we can
make sure the approximation quality is consistently bounded at every
point.

Second, the {\em seg-err} function checks the error introduced by
each segment, and captures the maximal error. That is, \[segerr(X,
\tilde{X})=\max_{s \in \tilde{X}}\{\sum_{i \in range(s)}
(x_i-\tilde{x}_i)^2\}\] Using the seg-err function, we can make sure
that the error introduced by every segment is bounded.

Using the two error measurement functions, we have two versions of
the error-bounded PLA problem.

\begin{problem}[PLA-PointBound problem]
Given an error-bound $\epsilon$, the {\em PLA-PointBound problem} is
to find an $\epsilon$-PLA $\tilde{X}$ such that $maxerr(X,
\tilde{X}) \leq \epsilon$ and $|\tilde{X}|$ is minimized. \boxend
\end{problem}

\begin{problem}[PLA-SegmentBound problem]
Given an error-bound $\epsilon$, the {\em PLA-SegmentBound problem}
is to find an $\epsilon$-PLA $\tilde{X}$ such that $segerr(X,
\tilde{X}) \leq \epsilon$ and $|\tilde{X}|$ is minimized. \boxend
\end{problem}

\nop{It is worth noting that different versions of the $\epsilon$-PLA
problem require different solutions. Generally, the PLA-PointBound
problem composes a stronger quality requirement than the
PLA-SegmentBound problem, and thus the PLA-PointBound problem is
more challenging.

A good solution to one problem does not necessarily result in a
good solution to the other. For instance, suppose that we have an
effective algorithm to solve the PLA-PointBound problem. One may
wonder whether we can translate the PLA-SegmentBound problem to the
PLA-PointBound problem by setting the error-bound (of each point) to
$\sqrt{\frac{\epsilon}{n}}$, where $n$ is the number of points
approximated by a line segment.

Unfortunately, it is hard to obtain the number of points in a
segment. The segments may have different lengths. Moreover, a
segment in the PLA-SegmentBound problem might approximate most
points well, but permits several exceptional points to have large
errors. A na\"ive translation of the PLA-SegmentBound problem to the
PLA-PointBound problem using a uniform error-bound on all points
excludes such segments. Similarly, we cannot easily reduce the
PLA-PointBound problem to the PLA-SegmentBound problem.}

\subsection{Related Work}

Piecewise linear approximation (PLA) has been well investigated
in~\cite{Dou, keogh01online, keogh98enhanced, qu98supporting, Sha,
Wang}. The idea behind PLA comes from the fact that a sequence of
line segments can be used to represent the time series while
preserving a low approximation error.\nop{Since only two points are
enough to specify a line segment, PLA can play the role of a simple
compression algorithm and make the space, transmission and
computation complexity of time series more efficient.} Standard
linear regression technique is widely used in most existing
piecewise linear approximation algorithms to calculate a line
segment approximating the original data with the minimum mean
squared error. Many of them~\cite{Dunh86,Good94,keogh98enhanced,qu98supporting} 
target at solving the conventional PLA
problem and may not be applicable to streaming data.

Despite the substantial research efforts in PLA
techniques~\cite{Dunh86, Good94, keogh01online, keogh_amnestic,
keogh98enhanced, qu98supporting}, existing solutions are not
tailored for data streams over resource-constrained sensor networks.
They either require complex computation or have high cost in space.
To the best of our knowledge, there has no implementation of these
algorithms in realistic sensor device.

In~\cite{EnergyChong}, the authors use PLA to estimate a time
series. But the authors put unnecessary constraints on the algorithm, which
requires the endpoints come from the original dataset. On the whole,
their algorithm can run in $O(n^2log n)$ time complexity and takes
$O(n)$ space complexity.

In~\cite{keogh01online}, Keogh et al.\ give a comprehensive review
on the existing techniques for segmenting time series. They
categorize the solutions into three different groups, namely sliding
window methods, top-down methods, and bottom-up methods. They then
take advantage of both sliding window and bottom-up methods and
design a Sliding-Window-And-Bottom-up (SWAB) algorithm. The SWAB
algorithm uses a moving window to constrain a time period in
consideration.

In~\cite{keogh_amnestic}, an amnesic function is introduced to give
weights to different points in the time series. The PLA-SegmentBound
problem is discussed in the context of Unrestricted Window with
Absolute Amnesic (UAA) problem, but complete solutions to this
problem are not provided in~\cite{keogh_amnestic}.

A solution to the PLA-PointBound problem is addressed
in~\cite{OptimalPLA} with a different definition of point error
bound. The algorithm is claimed to be optimal, but the time
complexity is $O(n^3)$ where $n$ is the number of points in the time
series. Moreover, no performance evaluation of the solution is
presented in the paper.

In summary, although the error-bounded PLA problem has been
investigated before, the problem has not been studied
systematically. No solutions applicable to data streams have been
developed, let alone solutions for resource-constrained sensor
networks.
\nop{~\footnote{To better appreciate our contributions, we
suggest the reviewers try to design a PLA algorithm that (1) is
online, (2) creates minimum number of line segments, (3) runs in
linear time, and (4) uses constant memory space, before reading our
following solutions.}.}

\section{Online Algorithms}
\label{sec:online}

In this section, we develop two online algorithms for the
PLA-PointBound and the PLA-SegmentBound problems, respectively. The
two algorithms share the same framework.
\\
\subsection{The Framework}

The framework of our algorithms works in a greedy manner. When
$x_1$, the first point in the stream, arrives, we store $x_1$. When
$x_2$ arrives, we also store $x_2$ since $x_1$ and $x_2$ can be
compressed by a segment exactly. When $x_3$ arrives, we check
whether $x_3$ can be compressed together with $x_1$ and $x_2$ by a
line segment satisfying the error-bound requirement. If so, we store
$x_3$. Otherwise, we output a line segment compressing $x_1$ and
$x_2$, remove $x_1$ and $x_2$ from the main memory, and store $x_3$.

Generally, imagine we have a buffer in main memory storing points
$x_i, x_{i+1}, \ldots, x_j$ such that the points in the buffer can
be compressed by a line segment satisfying the error-bound
requirement. When a new point $x_{j+1}$ arrives, we check whether
$x_{j+1}$ can be compressed together with $x_i, \ldots, x_j$ by a
line segment satisfying the error-bound requirement. If so, we add
$x_{j+1}$ to the buffer and move on to the next point. Otherwise, we
output a segment compressing $x_1, \ldots, x_j$ satisfying the
error-bound requirement, and remove them from the buffer. $x_{j+1}$
is then stored in the buffer.

Although the framework is simple, there are two critical issues that
need to be solved carefully in order to make sure that the runtime
of the algorithms is linear with respect to the number of points in
the streams, and the space size needed by the algorithms is bounded
by a constant.

First, how can we store the information about the points we have
seen but have not compressed? In the worst case, there can be an
unlimited number of such points (e.g., a times series where all
points take the same value). How can we summarize them using only
constant size memory?

Second, how can we determine whether a newly arrived point can be
compressed together with the points already in the buffer that have
been seen but have not been compressed? Revisiting those points one
by one leads to the runtime quadratic with respect to the number of
such points. As explained before, there can be an unlimited number
of such points. The overall time complexity is quadratic if those
points are revisited one by one.

Our central idea to tackle the above two challenges is the
following. Instead of storing the points explicitly, we monitor the
range of all possible line segments that can be used to compress the
points that have been seen but have not been compressed in a concise
way. When a new point arrives, we can check whether the point can be
compressed using some line segment in the range. If so, it means
that the new point can be compressed together with the points
accumulated. We only need to adjust the range of the possible line
segments to make sure the new point is also compressed. If not, it
means that the new point cannot be compressed together with the
points accumulated. A segment should be output.

\nop{In the rest of this section, we develop the techniques to implement
the above idea for the PLA-PointBound problem and the
PLA-SegmentBound problem, respectively.}

\subsection{Solving the PLA-PointBound
Problem}\label{sec:algo-pointbound}

A segment $s=((i, y_i), (j, y_j))$ can also be represented by the
left endpoint $(i, y_i)$, the slope $m=\frac{y_j-y_i}{j-i}$, and the
index of the right endpoint $j$.

\begin{figure}[t]
\centerline{\includegraphics[width=65mm]{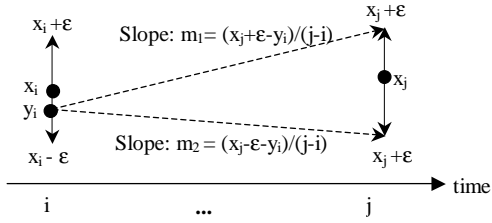}}
\caption{Ranges of possible line segments.}
\label{fig:OptimalPointBound}
\end{figure}

For two points $x_i$ and $x_j$ in a data stream, if a line segment
$s=((i, y_i), (j, y_j))$ with slope $m=\frac{y_j-y_i}{j-i}$ can
approximate $x_i$ and $x_j$, i.e., $|x_i-\tilde{x}_i| \leq \epsilon$
and $|x_j-\tilde{x}_j| \leq \epsilon$ where $\epsilon$ is the
error-bound, $s$ must satisfy the following four conditions.

\begin{equation} \label{eq1} (x_i - \epsilon) \le y_i \le (x_i + \epsilon) \end{equation}
\begin{equation} \label{eq2} m_1 = \frac{(x_j + \epsilon) - y_i}{j-i} \end{equation}
\begin{equation} \label{eq3} m_2 = \frac{(x_j - \epsilon) - y_i}{j-i} \end{equation}
\begin{equation} \label{eq4} m_2 \leq m \leq m_1 \end{equation}
Figure~\ref{fig:OptimalPointBound} illustrates the conditions and
their relations. Particularly, $m_1$ and $m_2$ are the slopes of the
two lines shown in the figure.

\begin{figure}[t]
\centerline{\includegraphics[width=60mm]{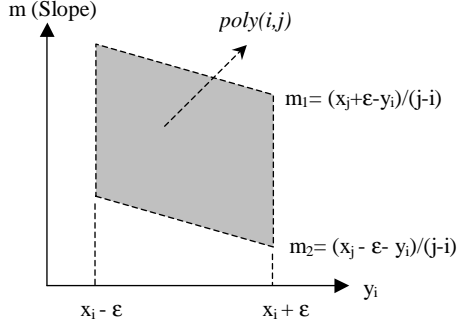}}
\caption{Polygon $poly(i, j)$.} \label{fig:Polygon}
\end{figure}

Since the line segments are determined by the value of the left
endpoint $y_i$ and slope $m$, we examine the distribution of points
$(y_i, m)$ that satisfy Equations~\ref{eq1} to~\ref{eq4}. As
illustrated in Figure~\ref{fig:Polygon}, the possible line segments
form a polygon $poly(i,j)$. We have the following important result.

\begin{lemma}[PLA-PointBound]\label{lem:pointbound}
A line segment of left endpoint $y_i$ and slope $m$ can approximate
points $x_i, \ldots, x_j$ with max-err at most $\epsilon$ if and
only if $(y_i, m)$ is in polygon $poly(i, i+1) \cap poly (i, i+2)
\cap \cdots \cap poly (i, j)$. \em

{\noindent\bf Proof.} The necessity follows with the definition of
$poly(i, j)$. For any line segment $s \not\in poly(i, i+1) \cap poly
(i, i+2) \cap \cdots \cap poly (i, j)$, there exists an index $k$
$(i \leq k \leq j)$ such that $s \not \in poly(i, k)$, i.e., $s$
cannot approximate either $x_i$ or $x_k$.

We prove the sufficiency by contradiction. Suppose a segment $s \in
poly(i, i+1) \cap poly (i, i+2) \cap \cdots \cap poly (i, j)$ but
$s$ cannot approximate $x_k$ $(i \leq k \leq j)$. Two situations may
arise. First, $k=i$. Then, $s \not\in poly(i, i+1)$ since $|x_i
-y_i| > \epsilon$ where $y_i$ is the value of $s$ on index $i$. Second,
$k \neq i$. Then, $s \not\in poly(i, k)$. In both cases, we have
contradictions. \boxend
\end{lemma}

\begin{figure}[t]
\setcounter{line}{0}
\begin{tabbing}
1234\=56\=78\=90\=01\=23\=45 \kill
  {\bf Input:} a data stream $X=x_1, x_2, \ldots$ and error-bound $\epsilon$;\\
  {\bf Output:} a list of line segments $\tilde{X}$ approximating $X$ \\
  \> such that $maxerr(X, \tilde{X})) \leq \epsilon$; \\
  {\bf Method:}\\
  \addtocounter{line}{1}\theline:
  \> $P=poly(1, 2)$; $i=1$; $j=3$; \\
  \addtocounter{line}{1}\theline:
  \> {\tt WHILE} (1) {\tt DO} $\{$ \\
  \addtocounter{line}{1}\theline:
  \>\> $P'= P \cap poly(i, j)$; \\
  \addtocounter{line}{1}\theline:
  \>\> {\tt IF} $P' \neq \emptyset$ {\tt THEN} $P=P'$, $j=j+1$; \\
  \addtocounter{line}{1}\theline:
  \>\> {\tt ELSE} $\{$ \\
  \addtocounter{line}{1}\theline:
  \>\>\> randomly choose a point $(y, m)$ in $P$; \\
  \>\>\>\textit{/*any point in $P$ meets the point error bound*/}\\
  \addtocounter{line}{1}\theline:
  \>\>\> output a line segment \\
  \>\>\>\> $((i, y), (j-1, y+(j-1-i)*m))$; \\
  \addtocounter{line}{1}\theline:
  \>\>\> $P=poly(j, j+1)$; $i=j$; $j=j+2$; \\
  \>\> $\}$ \\
  \> $\}$
\end{tabbing}
\hrule\caption{\label{fig:algo-pointbound} PointBound, an online algorithm for the PLA-PointBound problem.}
\end{figure}

Using Lemma~\ref{lem:pointbound}, we have algorithm PointBound, an
online algorithm as shown in Figure~\ref{fig:algo-pointbound}. We
maintain the intersection of polygons $poly(i, i+1)$, \ldots,
$poly(i, j)$, where $x_i$ is the first point that has not been
compressed yet in the data stream, and $x_j$ is the last point
arrived such that $poly(i, i+1) \cap \ldots \cap poly(i, j) \neq
\emptyset$.

When a new point $x_{j+1}$ arrives, we compute $poly(i, j+1)$ and
$poly(i, i+1) \cap \ldots \cap poly(i, j) \cap poly(i, j+1)$. If it
is $\emptyset$, then a line segment $s$ is randomly chosen to
approximate $x_i, \ldots, x_j$ such that $(y_i, m)$ is in $poly(i,
i+1) \cap \ldots \cap poly(i, j)$, where $y_i$ is the value of $s$
on index $i$, and $m$ is the slope of $s$. $s$ is output, and the
intersection of polygon is removed. $x_{j+1}$ and $x_{j+2}$ are used
to generate a new polygon $poly(j+1, j+2)$.

If $poly(i, i+1) \cap \ldots \cap poly(i, j) \cap poly(i, j+1) \neq
\emptyset$, then the intersection is kept, and the algorithm moves
on to the next point in the stream.

For any $i$ and $j$, $poly(i, j)$ is a parallelogram where there are
two edges parallel to the slope axis. It is easy to show that for
any $i$ and $j$, $\cap_{k=i}^j poly(i, k)$ is a convex polygon. In the worst
case, the edges of the intersection of parallelograms could be up to
$2(j-i+1)$, i.e., twice the number of parallelograms intersected. A
straightforward method keeping all edges of the intersection area
still has the quadratic time complexity and linear space complexity,
which are not applicable to data streams.

Fortunately, we do not need to record all edges of the intersection
polygon. Instead, {\em we need to record only up to $4$ edges to
determine whether a new point can be compressed together with the
points seen but not compressed.}

Using Equations~\ref{eq1} to~\ref{eq4}, it is easy to see that each
parallelogram has two properties: (1) Each parallelogram has two vertical edges and two sloping
edges with a negative slope value, as shown in
Figure~\ref{fig:Polygon}. The range of $y_i$ is the same for all
parallelograms (i.e., $x_i-\epsilon \le y_i \le x_i+\epsilon$). (2) For $j_2 > j_1 > i$, the absolute slope value of the two
sloping edges in $poly(i, j_2)$ is strictly smaller than the
absolute slope value of the two sloping edges in $poly(i, j_1)$.

Let us focus on the intersection
points of the upper sloping edge of parallelograms. The case for the
lower sloping edges can be analyzed similarly.

\begin{figure}[t]
\centerline{\includegraphics[width=85mm]{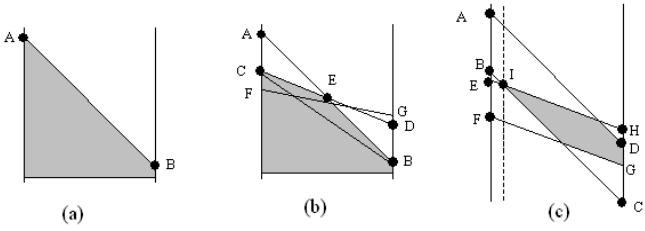}}
\caption{Using up to $4$ edges to represent the intersection
polygon.} \label{fig:PolygonIntersection}
\end{figure}

The situations are illustrated in
Figure~\ref{fig:PolygonIntersection}. Suppose that the first
parallelogram gives the upper sloping edge $AB$ with slope value
$m_{AB}$ as in Figure~\ref{fig:PolygonIntersection}(a). When a new
data point arrives, a new parallelogram is formed. In the worst
case, the upper sloping edge of the parallelogram $CD$ cuts $AB$
into two parts. Let $E$ be the intersection point between $AB$ and
$CD$, as shown in Figure~\ref{fig:PolygonIntersection}(b).

By the second property, we have $|m_{CD}|< |m_{AB}|$. Moreover, the
upper sloping edge $FG$ of any future parallelogram cannot cut both
$CE$ and $EB$ due to the smaller absolute slope value of $FG$ than
$m_{CD}$. In other words, if a future parallelogram intersects with
the current intersection polygon, the upper sloping edge of the
parallelogram can only cut either $CE$, $EB$ or the right vertical
edge. Instead of keeping $CE$ and $EB$, we can keep line segment
$CB$. Then, a future parallelogram intersects with the current
intersection polygon if and only if it cuts $CB$.

Generally, we only need to keep the line segment connecting the
left-most upper corner and the right-most upper corner for the upper
sloping edges. Similarly, we only need to keep the line segment
connecting the left-most lower corner and the right-most lower
corner for the lower sloping edges.

In addition to this two line segments, we need to keep the two
vertical edges in the intersection polygon. The reason is that the
intersection of two parallelograms may shrink the range of the
intersection, as illustrated in
Figure~\ref{fig:PolygonIntersection}(c), where parallelogram $ABCD$
intersects with parallelogram $EFGH$. The left vertical edge is
shrunk into a point $I$ right to the original edge.

In summary, we need to record only up to $4$ edges to determine
whether a new point can be compressed together with the points seen
but not compressed. This immediately leads to the following result.

\begin{theorem}[Complexity -- PointBound]\textbf{ }\\
The algorithm PointBound for the PLA-PointBound problem has the time
complexity $O(n)$ and the space complexity $O(1)$, where $n$ is the
number of points in a time series to be compressed. \boxend
\end{theorem}

Since algorithm PointBound only looks ahead for one point in the
data stream to output a line segment whenever necessary in the piecewise linear
approximation, it is an online algorithm and can be applied on data
streams.

\subsection{Solving the PLA-SegmentBound Problem}
\label{sec:onlineSegment}

We first present the following useful observation, to which a
similar result has been reported in~\cite{qu98supporting} without
proof.

\begin{lemma}\label{lem:segmentbound}
Suppose that a line segment $s$ approximates a fragment $X$ of $n$ points
$x_1, \ldots, x_n$ in a time series. Then, $s$ minimizes $segerr(s,
X)$ if the slope of $s$ is \begin{equation}\label{eq:seg3}m =
\frac{(\sum_{i = 1}^{n}ix_{i}) - \frac{1}{n}\sum_{i =
1}^{n}i\sum_{i = 1}^{n}x_{i}}{(\sum_{i = 1}^{n}i^{2})
-\frac{1}{n}(\sum_{i = 1}^{n}i)^{2}}\end{equation} and the left
endpoint of $s$ has value
\[m+\frac{\sum_{i=1}^n (x_i - i \cdot m)}{n}\] \em

{\noindent\bf Proof.} Consider a line segment $s$ approximating
fragment $X$. Let the left endpoint of $s$ be $(1, y_1)$ and the
slope be $m$. For each point $x_i$ $(1 \leq i \leq n)$, the error is
$|x_i-\tilde{x}_i|=|x_i-y_1-m(i-1)|$. Thus,
\begin{equation}
\begin{split}
segerr &= \sum_{i = 1}^{n}(x_i-y_1-m(i-1))^2 \label{eq:seg}
\end{split}
\end{equation}
Clearly, when $y_1=m+\frac{\sum_{i=1}^n (x_i - i \cdot m)}{n}$,
$segerr$ reaches the minimum value
\begin{equation}
\begin{split}
segerr=\sum_{i = 1}^{n} x_{i}^{2} + m^{2} \sum_{i = 1}^{n} i^{2} -
2m\sum_{i = 1}^{n} x_{i}i - \\ \frac{(\sum_{i = 1}^{n}(x_{i} - i*
m))^2}{n} \label{eq:seg2}
\end{split}
\end{equation}
From Equation~(\ref{eq:seg2}), when \[m = \frac{(\sum_{i =
1}^{n}ix_{i}) - \frac{1}{n}\sum_{i = 1}^{n}i\sum_{i =
1}^{n}x_{i}}{(\sum_{i = 1}^{n}i^{2}) -\frac{1}{n}(\sum_{i =
1}^{n}i)^{2}}\] $segerr$ is minimized. \boxend
\end{lemma}

\begin{figure}[t]
\setcounter{line}{0}
\begin{tabbing}
1234\=56\=78\=90\=01\=23\=45 \kill
  {\bf Input:} a data stream $X=x_1, x_2, \ldots$ and error-bound $\epsilon$;\\
  {\bf Output:} a list of line segments $\tilde{X}$ approximating $X$ \\
  \> such that $maxerr(X, \tilde{X})) \leq \epsilon$; \\
  {\bf Method:}\\
  \addtocounter{line}{1}\theline:
  \> $i=1$; $j=3$ \\
  \addtocounter{line}{1}\theline:
  \> $s=$ the line segment $((1, x_1), (2, x_2))$; \\
  \addtocounter{line}{1}\theline:
  \> {\tt WHILE} (1) {\tt DO} $\{$ \\
  \addtocounter{line}{1}\theline:
  \>\> $s'=$ the line segment identified in
  Lemma~\ref{lem:segmentbound} to \\
  \>\> compress $x_i, \ldots, x_j$; \\
  \addtocounter{line}{1}\theline:
  \>\> {\tt IF} $segerr(s', x_i \cdots x_j) \leq \epsilon$ {\tt THEN}
  \\
  \addtocounter{line}{1}\theline:
  \>\>\> $s=s'$; $j=j+1$; \\
  \addtocounter{line}{1}\theline:
  \>\> {\tt ELSE} $\{$\\
  \addtocounter{line}{1}\theline:
  \>\>\> output $s$; \\
  \addtocounter{line}{1}\theline:
  \>\>\> $i=j$; $j=j+2$; \\
  \addtocounter{line}{1}\theline:
  \>\>\> $s=$ the line segment $((i, x_i), (i+1, x_{i+1}))$; \\
  \>\> $\}$\\
  \> $\}$
\end{tabbing}
\hrule\caption{\label{fig:algo-segmentbound}SegmentBound, an online algorithm for the PLA-SegmentBound problem.}
\end{figure}

Lemma~\ref{lem:segmentbound} leads to algorithm SegmentBound, an
online algorithm for the PLA-SegmentBound problem as shown in
Figure~\ref{fig:algo-segmentbound}. Suppose $x_1, \ldots, x_n$ are
the points that have not been compressed yet. When a new point
$x_{n+1}$ arrives, we check whether the line segment identified by
Lemma~\ref{lem:segmentbound} can achieve the segment error bound. If
so, then $x_{n+1}$ is added into the buffer, and the algorithm moves
on to the next point in the stream. Otherwise, the line segment
suggested by Lemma~\ref{lem:segmentbound} for points $x_1, \ldots,
x_n$ is output, and $x_1, \ldots, x_n$ are considered compressed.
$x_{i+n}$ is added into the buffer.

\nop{The time cost is constant to apply Lemma~\ref{lem:segmentbound} to
check whether a newly arrived point can be compressed together with
the points that have been seen but have not been compressed. }
When a
new data point $x_{n+1}$ arrives, the left endpoint and the slope of
the line segment suggested by Lemma~\ref{lem:segmentbound} can be
calculated quickly. Technically, Equations~(\ref{eq:seg3}) and ~(\ref{eq:seg2}) indicate
that we need to calculate $\sum_{i = 1}^{n+1}i$, $\sum_{i =
1}^{n+1}x_{i}$, $\sum_{i = 1}^{n+1}x_{i}i$, $\sum_{i = 1}^{n+1}x_{i}^2$, and $ \sum_{i =
1}^{n+1}i^{2}$. Since we already have $\sum_{i = 1}^{n}i$,$\sum_{i = 1}^{n}x_{i}$, $\sum_{i = 1}^{n}x_{i}i$,$\sum_{i = 1}^{n}x_{i}^2$,
and $ \sum_{i =1}^{n}i^{2}$, \textit{the addition of the new point only incurs a constant
cost to update the values of $m$ and the left endpoint}. This leads to the following result.

\nop{Since each point in the data stream is processed in constant time,
and we only maintain the values of $\sum_{i = 1}^{n+1}i$, $\sum_{i =
1}^{n+1}x_{i}$, $\sum_{i = 1}^{n+1}x_{i}i$,$\sum_{i = 1}^{n+1   }x_{i}^2$, and $ \sum_{i =
1}^{n+1}i^{2}$, we have the following claim.}

\begin{theorem}[Complexity -- SegmentBound]\textbf{ }\\
The algorithm
SegmentBound for the PLA-SegmentBound problem has the time
complexity $O(n)$ and space complexity $O(1)$, where $n$ is the
number of points in a time series to be compressed. \boxend
\end{theorem}
\nop{ Similar to algorithm PointBound, algorithm SegmentBound only
looks ahead for one point in the data stream to output a line
segment whenever necessary in the piecewise linear approximation.
Thus, it is an online algorithm and can be applied on data streams.}

\section{Optimality}
\label{sec:optimal}

\nop{In this section, we address the quality of algorithms
PointBound and SegmentBound. Recall that, as defined in
Section~\ref{sec:prob}, in the error-bounded PLA problem, we want to
minimize the number of segments used to compress a time series.}

\begin{theorem}[PLA-PointBound quality]\label{thm:optimal-point}\textbf{ }\\
The PointBound algorithm in Section~\ref{sec:algo-pointbound}
produces a minimum number of segments to compress a time series. \em

{\noindent\bf Proof.} For a time series $X=x_1, \ldots, x_n$, let
$l=\min\{|\tilde{X}|\}$, where $\tilde{X}$ is an $\epsilon$-PLA
approximating $X$ (i.e., $maxerr(X, \tilde{X}) \leq \epsilon$). We
conduct an induction on $l$ to show that algorithm PointBound
outputs an $\epsilon$-PLA of $l$ line segments.

(Base case) Consider $l=1$, i.e., there exists a line segment that
approximates the whole time series. According to
Lemma~\ref{lem:pointbound}, $poly(1, 2) \cap \cdots \cap poly(1, n)
\neq \emptyset$. Thus, algorithm PointBound finds a line segment $s$
approximating $x_1, \ldots, x_n$ and $maxerr(s, X) \leq \epsilon$.

(Induction) Assume that, when $l \leq k$, algorithm PointBound finds
an $\epsilon$-PLA $\tilde{X}$ of $l$ line segments to approximate
$X$. Now, let us consider the case of $l=(k+1)$, i.e., there exists
an optimal $\epsilon$-PLA $\tilde{Y}=\{s_1, \ldots, s_{k+1}\}$ that
approximates $X$.

Suppose that $s_1$ approximates $x_1, \ldots, x_m$. Let us assume that
$s_1'$ output by algorithm PointBound approximates points $x_1,
\ldots, x_{m'}$. Due to Lemma~\ref{lem:pointbound}, $poly(1, 2) \cap
\cdots \cap poly(1, m) \neq \emptyset$. Thus, $s_1'$ must
approximate $x_1, \ldots, x_m$ with the quality guarantee, i.e.,
$maxerr(s_1', x_1 \cdots x_m) \leq \epsilon$. In other words, $m'
\geq m$.

If $m=m'$, then points $x_{m+1}, \ldots, x_n$ in $X$ can be
approximated by an $\epsilon$-PLA of $(l-1) = k$ line segments.
According to the assumption, algorithm PointBound finds an
$\epsilon$-PLA of $(l-1)$ line segments approximating $x_{m+1},
\ldots, x_n$.

Suppose that $m' > m$. Since $x_{m+1}, \ldots, x_n$ can be approximated
by an $\epsilon$-PLA of $(l-1)$ line segments, a proper subset
$x_{m'+1}, \ldots, x_n$ must also be approximated by an
$\epsilon$-PLA of at most $(l-1)=k$ line segments. We only need to
drop the segments approximating $x_{m+1}, \ldots, x_{m'}$. According
to the assumption, algorithm PointBound finds an $\epsilon$-PLA of
the minimum number of line segments to approximate points $x_{m'+1},
\ldots, x_n$.

In summary, algorithm PointBound finds an $\epsilon$-PLA of
$l=(k+1)$ line segments approximating $X$. \boxend
\end{theorem}

Similarly, we can also show the optimality of the SegmentBound
algorithm. \nop{Limited by space, we omit the details here.}

\begin{theorem}[PLA-SegmentBound quality]\textbf{ }\\
The SegmentBound algorithm in Section~\ref{sec:onlineSegment}
produces a minimum number of segments to compress a time series. \em
\end{theorem}

Although the number of line segments used to approximate a time
series is a good measure on the compression quality, it is not
directly translated to compression ratio. For example, in our
methods, the endpoints of segments are not constrained. Thus, two
points are needed to represent a segment. On the other hand, a PLA
using connecting segments (i.e., two consecutive segments share the
same endpoint) may use more segments but achieve a better
compression ratio since only one point is needed to represent a
segment except for the first segment.

\nop{How good compression can algorithms PointBound and SegmentBound
achieve? We have the following result.}

\begin{theorem}[Compression factor]\textbf{ }\\
Algorithms PointBound and SegmentBound have an
approximation factor of $2$ to the optimum compression factor that
an $\epsilon$-PLA can achieve. \em

{\noindent\bf Proof.} We only show the case for the PointBound
algorithm. The same argument applies to the SegmentBound algorithm.

For any time series $X$ of $m$ points, suppose that the PointBound
algorithm approximates $X$ using $n$ line segments. Then, according
to Theorem~\ref{thm:optimal-point}, any PLA cannot have less than
$n$ line segments. To represent $n$ line segments, at least $(n+1)$
points are needed. Thus, the optimum compression ratio using PLA is
at most $\alpha_{opt}=\frac{m}{n+1}$.

The line segments generated by the PointBound algorithm may not be
connecting. Thus, at most $2n$ points are needed to represent the
$n$ line segments. The worst case compression ratio of the
PointBound algorithm is $\alpha_{PointBound}=\frac{m}{2n}$. Clearly,
$\frac{\alpha_{opt}}{\alpha_{PointBound}}=\frac{2n}{n+1}<2$. \boxend
\end{theorem}

\section{PLAZA for Tiny Sensors}
\label{sec:PLAZA}

Although algorithm PointBound is optimal for the PLA-PointBound
problem, it still may be too computation intensive for tiny,
resource-constrained sensors due to two reasons.

First, algorithm PointBound may generate non-connecting segments
such that each segment requires the transmission of two endpoints.
As analyzed before, connecting line segments reduce the data
transmission volume since each segment (except the first one)
requires the transmission of only one endpoint. Second, algorithm
PointBound has to calculate intersection of parallelograms. The
computation may be too heavy for tiny, resource-constrained
sensor nodes.

In this section, we design a simple, fast online algorithm PLAZA
(\underline{P}iecewise \underline{L}inear \underline{A}pproximation
with \underline{Z}oning \underline{A}ngle) for the PLA-PointBound
problem. PLAZA generates connecting line segments. Although PLAZA is
not optimal in the number of line segments used for
approximation, it is light in computation and very effective in
compression ratio, as will be verified by our experiments.

\subsection{PLAZA}

PLAZA builds on the concept of zoning angle. Given an error bound
$\epsilon$ and two points $(i,x_i)$ and $(k,x_k)$ ($i<k$), the {\em
zoning angle} from $(i,x_i)$ to $(k,x_k)$, denoted by
$\theta_{(i,k)}^\epsilon$, is defined as the angle that has
$(i,x_i)$ as the endpoint, $((i,x_i),(k,x_k))$ as the bisector, and
has a degree of $2\arctan\frac{\epsilon}{|x_ix_k|}$, where
$|x_ix_k|=\sqrt{(k-i)^2+(x_k-x_i)^2}$.

\begin{figure}[t]
\centerline {\epsfxsize = 3.3 in \epsfbox{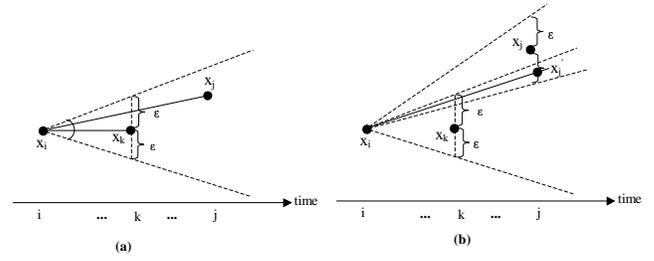}}
\caption{An example of zoning angle}
\label{fig:feasible}
\end{figure}

Figure~\ref{fig:feasible}(a) shows an example of zoning angle
$\theta_{(i,k)}^\epsilon$. The zoning angle defines a zone to include any potential line segments that can be used to compress $x_i$ and $x_k$.

We observe the following important results. Their proof is trival and is omitted due to space limit.

\begin{lemma}\label{lem:zoning}
For three points $x_i, x_k, x_j (i<k<j)$ in a time series, the line
segment $((i,x_i), (j,x_j))$ approximates $x_k$ with error up to
$\epsilon$ if and only if the line segment $((i,x_i),(j,x_j))$ falls
in the zoning angle $\theta_{(i,k)}^\epsilon$. \em
\nop{
{\noindent\bf Proof:} (Necessity) Without loss of generality, we
assume that $(j,x_j)$ is above the line $((i,x_i),(k,x_k))$, as
shown in Figure~\ref{fig:feasible}(a). If $((i,x_i),(j,x_j))$
approximates $x_k$ with error up to $\epsilon$, the vertical
distance between point $(k,x_k)$ and line $((i,x_i),(j,x_j))$ must
be smaller than $\epsilon$. That is, the line segment
$((i,x_i),(j,x_j))$ must fall in the zoning angle
$\theta_{(i,k)}^\epsilon$.

(Sufficiency) If the line segment $((i,x_i),(j,x_j))$ falls in the
zoning angle $\theta_{(i,k)}^\epsilon$, the degree of angle $\angle
x_jx_ix_k$ is smaller than $\arctan\frac{\epsilon}{|x_ix_k|}$.
Therefore, the vertical distance between point $(k,x_k)$ and line
$((i,x_i),(j,x_j))$ is smaller than $\epsilon$, which means, line
segment $((i,x_i),(j,x_j))$ approximates $x_k$ with error up to
$\epsilon$. \boxend}
\end{lemma}

\begin{lemma}\label{lem:overlap}
For three points $x_i, x_k, x_j (i<k<j)$ in a time series, if zoning
angle $\theta_{(i,j)}^\epsilon$ has no overlap with zoning angle
$\theta_{(i,k)}^\epsilon$, there does not exist a line segment $s$
with $(i,x_i)$ as the left endpoint such that $maxerr(s, x_i \cdots
x_k \cdots x_j) \leq \epsilon$.  \em
\nop{
{\noindent\bf Proof:} We prove by contradiction. Assume that there
is a line segment $s$ with $(i,x_i)$ as the left endpoint such that
$s$ approximates both $x_k$ and $x_j$ with an error up to
$\epsilon$. Based on Lemma~\ref{lem:zoning}, this line segment must
fall in both $\theta_{(i,k)}^\epsilon$ and
$\theta_{(i,j)}^\epsilon$, which means $\theta_{(i,k)}^\epsilon$ and
$\theta_{(i,j)}^\epsilon$ overlap. Contradiction. \boxend.}
\end{lemma}

\begin{figure}[t]
\setcounter{line}{0}
\begin{tabbing}
1234\=56\=78\=90\=01\=23\=45 \kill
  {\bf Input:} a data stream $X=x_1, x_2, \ldots$ and error-bound $\epsilon$;\\
  {\bf Output:} an $\epsilon$-PLA $\tilde{X}$ of a list of {\em connecting} line \\
  \> segments, i.e., $maxerr(X, \tilde{X})) \leq \epsilon$; \\
  {\bf Method:}\\
  \addtocounter{line}{1}\theline:
  \> $i=1$; $angle =\theta_{(1,2)}^\epsilon$; \\
  \addtocounter{line}{1}\theline:
  \> $s=$ line segment $((1, x_1), (2, x_2))$; $j=3$;\\
  \addtocounter{line}{1}\theline:
  \> {\tt WHILE} (1) {\tt DO} $\{$ \\
  \addtocounter{line}{1}\theline:
  \>\> $angle= angle \cap \theta_{(i,j)}^\epsilon$; \\
  \addtocounter{line}{1}\theline:
  \>\> {\tt IF} $angle \neq 0$ {\tt THEN} $\{$\\
  \addtocounter{line}{1}\theline:
  \>\>\> {\tt IF} segment $((i,x_i), (j,x_j))$ falls in $angle$ \\
  \addtocounter{line}{1}\theline:
  \>\>\> {\tt THEN} $s=$ line segment $((i, x_i), (j, x_j))$;\\
  \addtocounter{line}{1}\theline:
  \>\>\> {\tt ELSE} $\{$ \\
  \addtocounter{line}{1}\theline:
  \>\>\>\> $x_j^{'}$ = the value of the bisector line of \\
  \>\>\>\>  $angle$ at index $j$ as shown in Figure~\ref{fig:feasible}(b); \\
  \addtocounter{line}{1}\theline:
  \>\>\>\> $s=$ the line segment $((i, x_i), (j, x_j^{'}))$;\\
  \addtocounter{line}{1}\theline:
  \>\>\>\> $x_j = x_j^{'}$;\\
   \addtocounter{line}{1}\theline:
  \>\>\> $\}$ \\
  \addtocounter{line}{1}\theline:
  \>\>\> $j=j+1$;\\
 \addtocounter{line}{1}\theline:
  \>\> $\}$\\
  \addtocounter{line}{1}\theline:
  \>\> {\tt ELSE} $\{$\\
  \addtocounter{line}{1}\theline:
  \>\>\> output $s$;\\
  \addtocounter{line}{1}\theline:
  \>\>\> $i=j-1$; $x_i=x_{j-1}$; $j=j+1$;\\
  \addtocounter{line}{1}\theline:
  \>\>\> $angle =\theta_{(i,i+1)}^\epsilon$; \\
   \addtocounter{line}{1}\theline:
  \>\>\> $s=$ line segment $((i, x_i), (i+1, x_{i+1}))$; \\
  \addtocounter{line}{1}\theline:
  \>\> $\}$\\
   \addtocounter{line}{1}\theline:
  \> $\}$
\end{tabbing}
\hrule\caption{\label{fig:alg} Algorithm PLAZA.}
\end{figure}

Algorithm PLAZA works as follows. Starting from a point $x_i$,
Lemma~\ref{lem:zoning} is used to check if there is a line segment
approximating points between indexes $i$ and $j (i<j)$. Moreover,
Lemma~\ref{lem:overlap} is used to check if searching further in the
time series is futile. The pseudocode of PLAZA is shown in
Figure~\ref{fig:alg}. Algorithm PLAZA scans each point in a data
stream only once and stores only the zoning angle and the current
approximating segment in main memory, the algorithm clearly has
linear time complexity and constant space complexity.

\nop{Compared to algorithm PointBound, PLAZA requires much simpler
computation. The major operation is to construct the zoning angle
and to calculate the intersection of zoning angles. The exact degree
of each angle is not required. Instead, we only need to record the
endpoint and the two edges of the angle. Similar to algorithm
PointBound, PLAZA returns endpoints of segments that may not
necessarily belong to the original time series (i.e., the value of
$x_j$ may be changed as shown in line $11$ in the pseudo code).}

\subsection{Benchmarking PLAZA}

PLAZA creates connecting line segments. Only transmission of one
point is needed for each line segment except for the first line
segment. This feature distinguishes PLAZA from algorithms PointBound
and SegmentBound. What is the optimal compression that can be
achieved by an $\epsilon$-PLA consisting of only connecting line
segments?

The idea behind the optimal PLAZA benchmark algorithm is similar to
that of algorithm PointBound. The main difference is that, unlike
the PointBound algorithm, we do not start the new segment with the
initial condition $x_{i} - \epsilon \le y_i \le x_{i} + \epsilon$,
where $y_i$ is the value of the left endpoint of the new segment.
Instead we set a smaller range on $y_i$ to guarantee the
connectivity of two consecutive segments. Specifically, to decide
the range of $y_i$, we use the last non-empty polygon intersection
in the previous point.

\nop{
Suppose points $x_1, \ldots, x_{j-1}$ are checked but have not been
compressed yet, i.e., $poly(1,2) \cap \cdots \cap poly(1,j-1) \neq
\emptyset$. We keep the exact instersection of the parallelograms.
The intersection is used to confine the range of $y_j$. Apparently,
as the intersection is a convex polygon, only the corners of the
polygon need to be checked when looking for the minimum and maximum
values. This fact helps the search of $y_j$.}

We find the optimal solution by a thorough search. Starting from
$x_1$, we try all values of $j$ such that $x_1, \ldots, x_j$ can be
approximated by a line segment with maximal error $\epsilon$. For
each such a subset $x_1, \ldots, x_j$, we compute the intersection
of parallelograms $poly(1,2) \cap \cdots \cap poly(1,j)$, and try to
find a line segment with left endpoint $(j, y_j)$ that can
approximate some points $x_{j+1}, \ldots, x_i$ where $j+1 < i$ and
$y_j$ is in the range confined by $poly(1,2) \cap \cdots \cap
poly(1,j)$. By doing so, the first and the second line segments are
connected. We conduct a depth-first search to find an $\epsilon$-PLA
consisting of the minimum number of connecting line segments.
Limited by space, we omit the details here.

The optimal PLAZA benchmark is an offline algorithm: it assumes the
time series is given and can be scanned multiple times. Its
complexity is far above linear due to the thorough search. This
algorithm is obviously not suitable for online compression of data
streams. It is for comparison purpose only.

\section{Experimental Evaluation}
\label{sec:exp}

In this section, we evaluate the performance of our online
algorithms by simulation in Matlab and by real implementation with
MICA2 motes~\cite{MICA2}.

\nop{
\subsection{Existing Work}

The work in \cite{keogh01online, keogh_amnestic} is related to our online
algorithms. After much deliberation, however, we find it is
impossible to get fair comparison between our online algorithms and
those methods due to the following reasons.

First, the SWAB algorithm in~\cite{keogh01online} uses a moving
window to constrain a time period in consideration. Its performance
largely depends on the size of the moving window. As such, SWAB
is largely different from our methods that do not maintain any window. Furthermore,
it is hard to allocate enough space for the moving window in
space-limited tiny sensors.

Second, the algorithms in~\cite{keogh_amnestic} use an amnesic
function to give weights to different points in the time series. The
focus and the application context are different from ours.

Due to the above consideration, in the rest of this section, we
focus on the implementation and the evaluation of our online
algorithms. }

\subsection{Experimental Setting}

We generated two audio files for test. The first file includes human
voice with the sampling rate of $8$ khz in mono channel. The second
file includes piano music with the sampling rate of $44$ khz in mono
channel. Each file includes $1,000,000$ samples, and the size of
each sample is $16$ bits. Figures~\ref{fig:HumanWave}
and~\ref{fig:MusicWave} show the waveform of the human voice data
and the waveform of the piano music, respectively. It can be seen
that the music data is much ``smoother" than the human voice data.
We use the files to test the performance of our online algorithms in
bandwidth saving. We measure two metrics:

\begin{figure*}[htb!]
\begin{center} \begin{minipage}{75 mm}
    \centerline{\includegraphics[width=75mm]{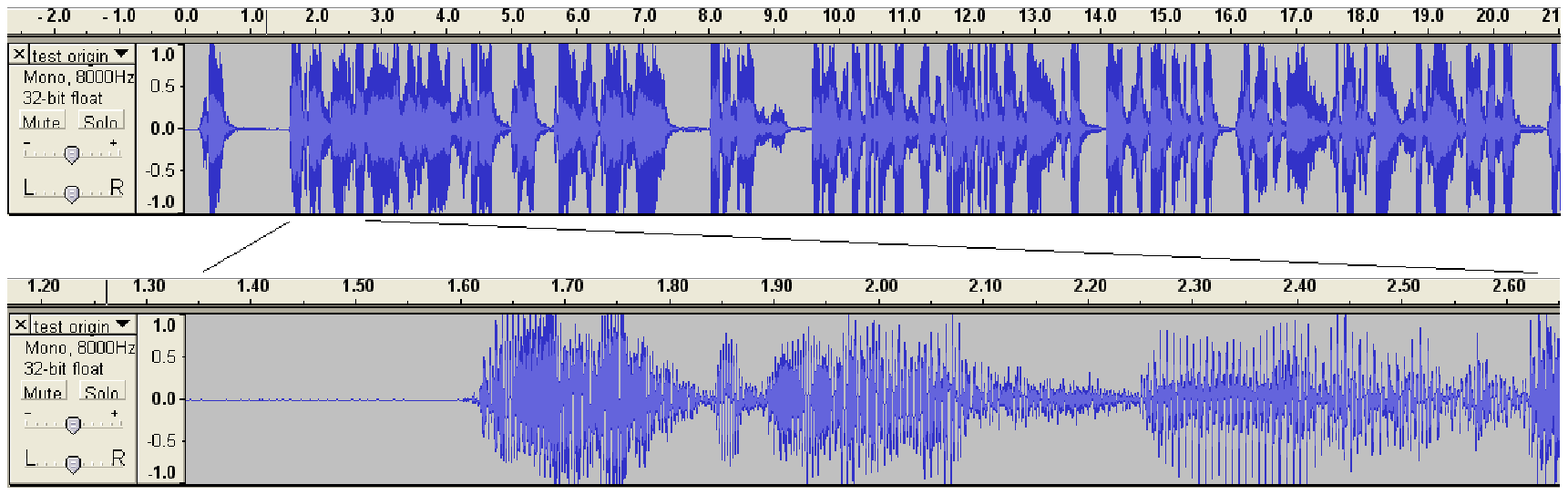}}
    \caption{\label{fig:HumanWave} The waveform of the human voice
    data (the lower part is in a smaller time scale).}
  \end{minipage}\hspace{2mm}
  \begin{minipage}{75 mm}
    \centerline{\includegraphics[width=75 mm]{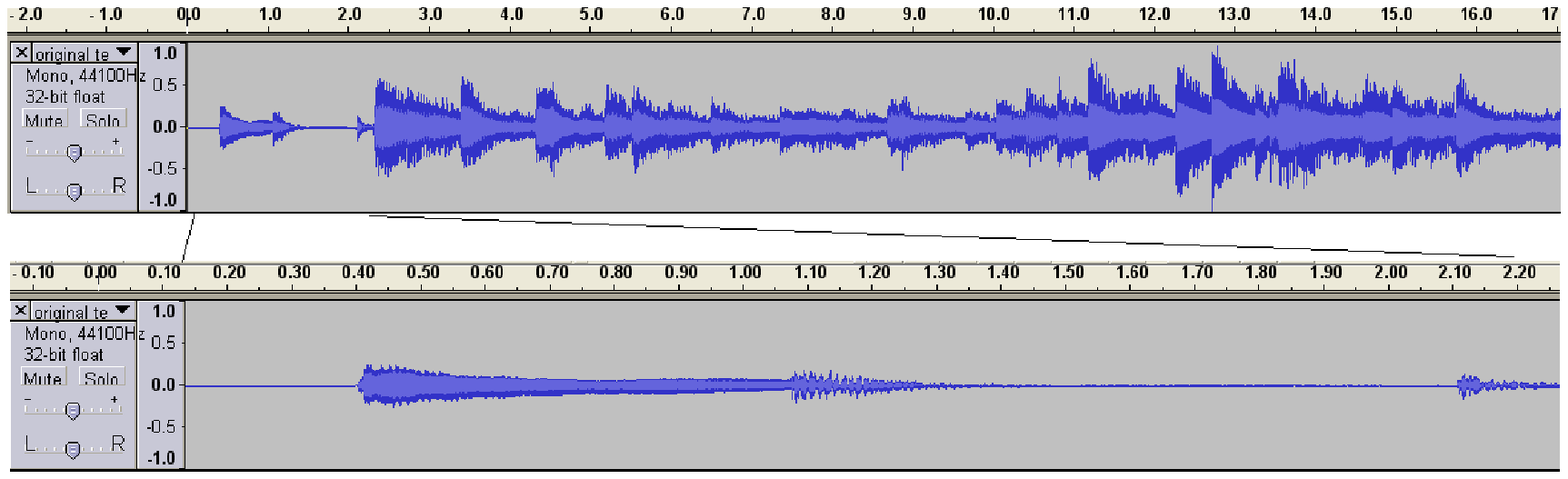}}
    \caption{\label{fig:MusicWave} The waveform of the piano music
    data (the lower part is in a smaller time scale).}
  \end{minipage}
\end{center}

 \begin{center}
  \begin{minipage}{75 mm}
    \centerline{\includegraphics[width=60mm]{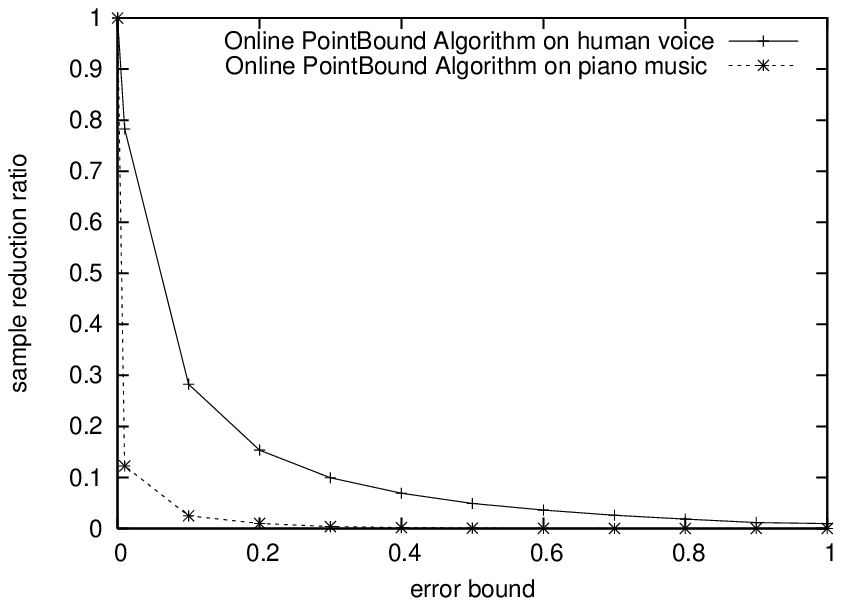}}
    \caption{\label{fig:PointBoundProblem} The sample reduction ratio
    of PointBound.}
  \end{minipage}\hspace{2mm}
  \begin{minipage}{75 mm}
    \centerline{\includegraphics[width=60mm]{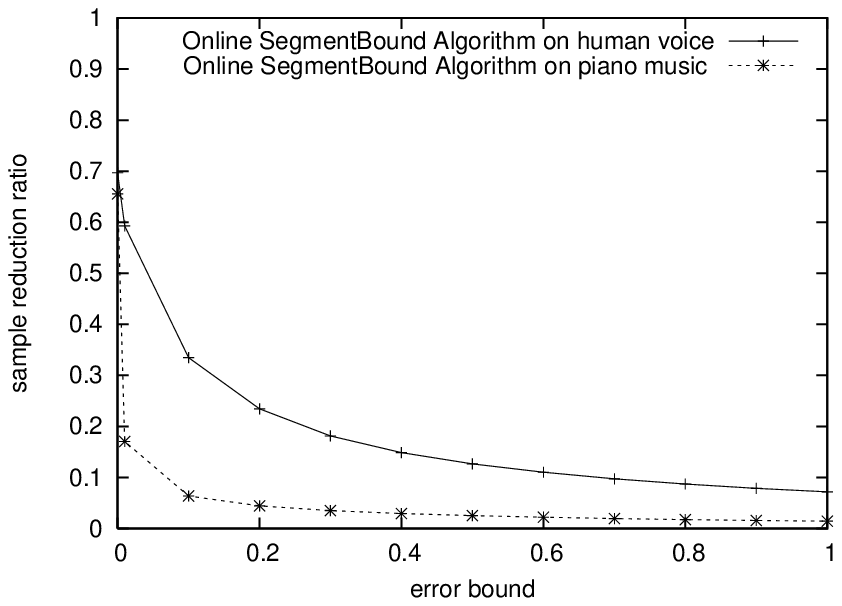}}
    \caption{\label{fig:SegmentBoundProblem} The sample reduction ratio
    of SegmentBound.}
  \end{minipage}
\end{center}
\end{figure*}

\begin{enumerate}

\item {\em Sample reduction ratio (inverted compression ratio)}.
It is defined as the total number of points to represent the
$\epsilon$-PLA divided by the total number of points in the original
time series.

\item {\em Distortion}. It is defined as
$\frac{\sum_{i=1}^n (x_i-\tilde{x_i})^2}{n}$, where $n$ is the total
number of points in the time series, $x_i$ is the original value,
and $\tilde{x_i}$ is the approximated value of $x_i$.

\end{enumerate}

In simulation, we apply the online algorithms on the audio files and
measure the sample reduction ratio. Simulation results are reported
in Section~\ref{sec:result1} and Section~\ref{sec:result2}. In the
test using MICA2 motes, the original audio files are played on a
desktop computer and are monitored and transmitted with a MICA2 mote
over wireless channel to a laptop computer. More details are
provided in Section~\ref{sec:implementation}.

\subsection{Results on Quality}

\nop{ In this section, we test the sample reduction ratio and
distortion using the two audio data sets.}

\label{sec:result1}
\subsubsection{Results on Sample Reduction Ratio}

Figures~\ref{fig:PointBoundProblem}
and~\ref{fig:SegmentBoundProblem} show the results of algorithms
PointBound and SegmentBound, respectively, with respect to various
error bound values. As shown in the figures, we can obtain a higher
bandwidth saving on piano music than on human voice. \nop{As shown in
Figures~\ref{fig:HumanWave} and~\ref{fig:MusicWave}, the waveform of
the piano music is much ``smoother'' than that of the human voice.}
By replaying the audio files recovered from the samples by our
algorithms, we perceive that the human voice recovered from the
samples by our algorithms is fully recognizable with the segment
error bound up to $0.4$, or with the point error bound up to $0.2$.
The quality of recovered piano music is acceptable to us with the
segment error bound up to $0.2$, or with the point error bound up to
$0.1$.

Figures~\ref{fig:PointBoundProblem} and~\ref{fig:SegmentBoundProblem}
 clearly demonstrate significant
bandwidth saving. With the online algorithms, we only need to
transmit around $5\%$ of the original sample size for piano music
and around $20\%$ of the original sample size for human voice. As
such, both sound files can be transmitted with the current sensor
nodes.

\begin{figure}[htb!]
  \centerline{\includegraphics[width=60mm]{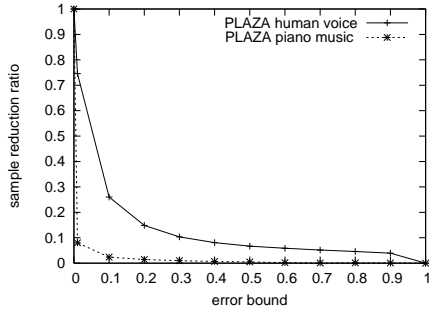}}
  \caption{\label{fig:PointBoundProblemPLAZA} The sample reduction ratio
  of PLAZA .}
\end{figure}

Figure~\ref{fig:PointBoundProblemPLAZA} shows the sample reduction
ratio of algorithm PLAZA with respect to various point error bounds.
We can observe the similar phenomenon as in
Figures~\ref{fig:PointBoundProblem} and ~\ref{fig:SegmentBoundProblem}. With PLAZA, we perceive that the recovered human voice is fully recognizable
with the (point) error bound up to $0.2$, and the quality of recovered
piano music is acceptable to us with the (point) error bound up to
$0.1$. From Figure~\ref{fig:PointBoundProblemPLAZA}, the above
qualities correspond to the bandwidth reduction of nearly $3\%$ of
the original data size for piano music and about $15\%$ of the
original data size for human voice.

One interesting phenomenon is that the SegmentBound algorithm can reduce sample transmission volume even
if the error bound is set to zero, as shown in
Figure~\ref{fig:SegmentBoundProblem}. This is because in the audio
files, there are some silent periods where the sample values are close to
zeros. The SegmentBound algorithm finds a line segment to approximate
those situations. This nice feature, however, does not exist in the
algorithms for the PLA-PointBound problem. If the error bound is
zero, the initial polygon is empty in the PointBound algorithm, and the
degree of the initial feasible angle is zero in PLAZA, resulting in no sample reduction.

\begin{figure}[htb!]
  \centerline{\includegraphics[width=60mm]{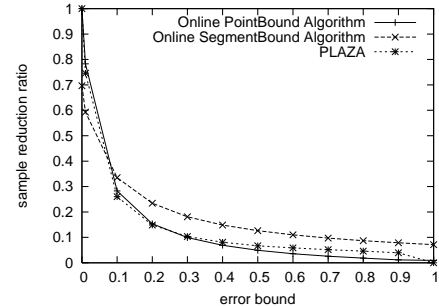}}
  \caption{\label{fig:HuVoice} Comparison of the three
  algorithms on the human voice data set.}
\end{figure}

Figure~\ref{fig:HuVoice} compares algorithms PLAZA, PointBound, and
SegmentBound on the human voice data set. The gap between algorithms
PLAZA and PointBound is very small when the error bound is less than
$0.5$. Algorithm PointBound leads to more samples than algorithm
PLAZA when the error bound is less than $0.3$. The gap between
algorithm SegmentBound and the two algorithms for the
PLA-PoinBound problem comes from the fact that, using the same error
bound value, the PLA-SegmentBound problem puts a tighter error
constraint than the PLA-PointBound problem. We observe the similar
performance comparison of the three algorithms on the piano data
set, but omit the figures here due to space limit.\\ \nop{We omit the details for the interest of space.}

\subsubsection{Results on Distortion}

\begin{figure*}[htb!]
\begin{center}
  \begin{minipage}{75 mm}
    \centerline{\includegraphics[width=60mm]{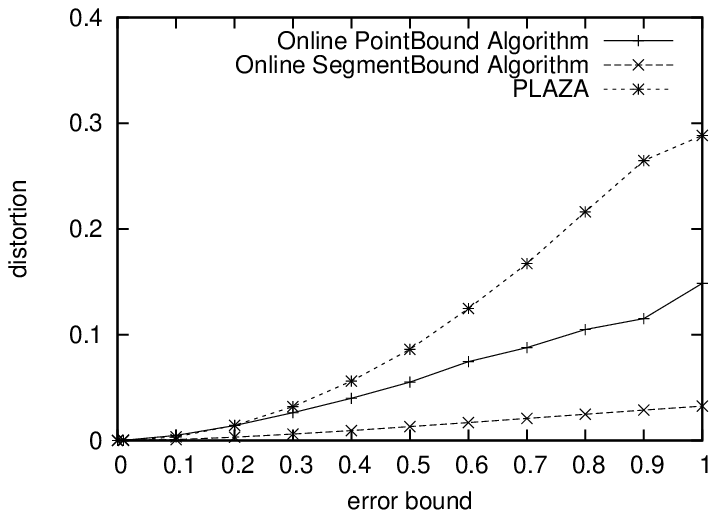}}
    \caption{\label{fig:Humandistortion} The distortion on the
    human voice dataset.}
  \end{minipage}\hspace{2mm}
  \begin{minipage}{75 mm}
    \centerline{\includegraphics[width=60mm]{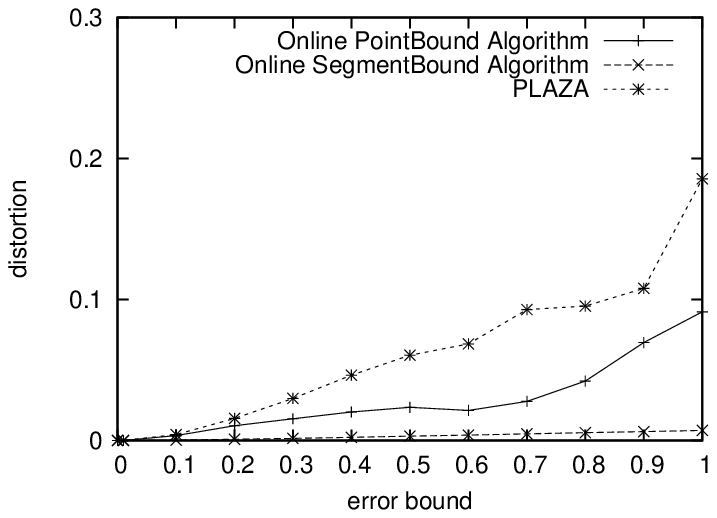}}
    \caption{\label{fig:musicdistortion} The distortion on the
    piano music voice dataset.}
  \end{minipage}
\end{center}
\end{figure*}

In Figures~\ref{fig:Humandistortion} and~\ref{fig:musicdistortion},
we quantitatively show the distortion of our algorithms on the human
voice data set and the piano music data set, respectively. The
overall distortion on human voice is larger than that on piano music
due to the ``smoother" waveform in the music data set. With the same
error bound, algorithm PLAZA has the largest distortion. Algorithm
PointBound is the next. Algorithm SegmentBound has the smallest
distortion because the same error bound on the PLA-SegmentBound
problem and the PLA-PointBound problem poses a tighter error
constraint on the PLA-SegmentBound problem. The smaller distortion,
however, comes with the cost of lower bandwidth saving as analyzed
before. \nop{This is the tradeoff between bandwidth saving and quality of
recovery.}

\subsection{Benchmarking PLAZA}
\label{sec:result2}

We test the performance of PLAZA comparing to the optimal solution
of its kind (i.e., using connecting line segments to tackle the
PLA-PointBound problem). Due to the high complexity of the PLAZA
Benchmark method, the audio files are too big to obtain the optimal
results within reasonable time. We have to use a small portion of
the audio files for this test.

Interestingly, the PLAZA method and the optimal PLAZA benchmark
algorithm generate very similar PLA line segments. Audio files are 
usually filled with short silent periods where sample values are close to $0$. Thus,
algorithm PLAZA can obtain line segments very similar to those
computed by the benchmark algorithm. We omit the detailed figures due to space limit.

\nop{
To further test algorithm PLAZA with more difficult scenarios, we
artificially generated some data sets containing Gaussian noise with
the mean equal to $0$ and the variance equal to $1$. We generate $5$
Gaussian noise data sets. Each data set has $1,000$ samples. We run
the simulation on the $5$ data sets and calculate the average sample
reduction ratio as the final result.

\begin{figure}[htb!]
  \centerline{\includegraphics[width=60mm]{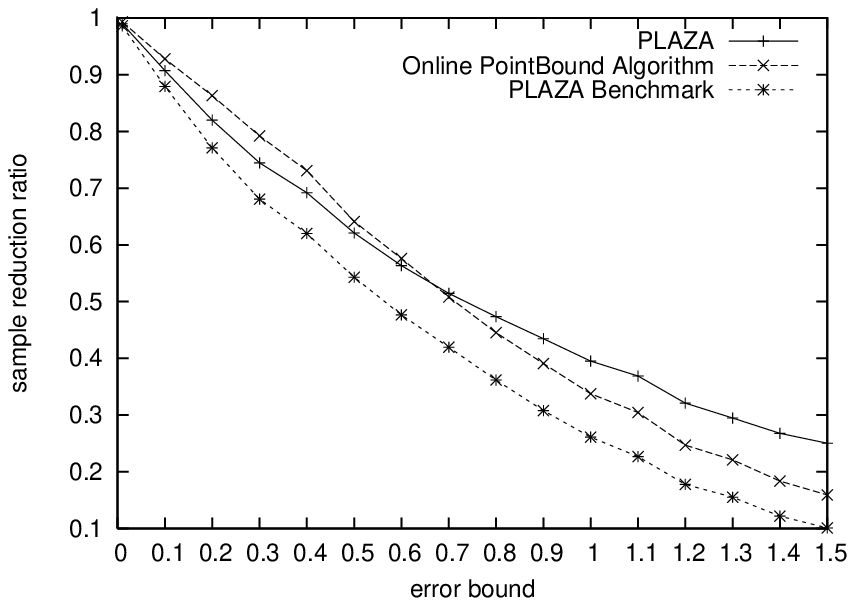}}
  \caption{\label{fig:PLAZAComparison} The comparison between algorithm
  PLAZA and the PLAZA benchmark method.}
\end{figure}

Figure~\ref{fig:PLAZAComparison} compares algorithms PLAZA,
PointBound, and the PLAZA benchmark. Since algorithm PointBound may
generate disconnected line segments, it has to send two endpoints
for each line segments. In contrast, algorithm PLAZA and the PLAZA
benchmark method always generate connecting segments, where every
segment, except for the first one, requires the transmission of only
one endpoint. The figure shows that algorithm PointBound actually
needs more endpoints than algorithm PLAZA when the error bound is
small, even though algorithm PointBound is optimal in terms of the
number of segments. Compared to the PLAZA benchmark algorithm,
algorithm PLAZA always generates more samples. The gap between
algorithm PLAZA and the PLAZA benchmark method, however, is not
significant when the error bound is small.
}

\subsection{Results on Real Sensors}
\label{sec:implementation}

\begin{figure}[t!]
  \centerline{\includegraphics[width=60mm]{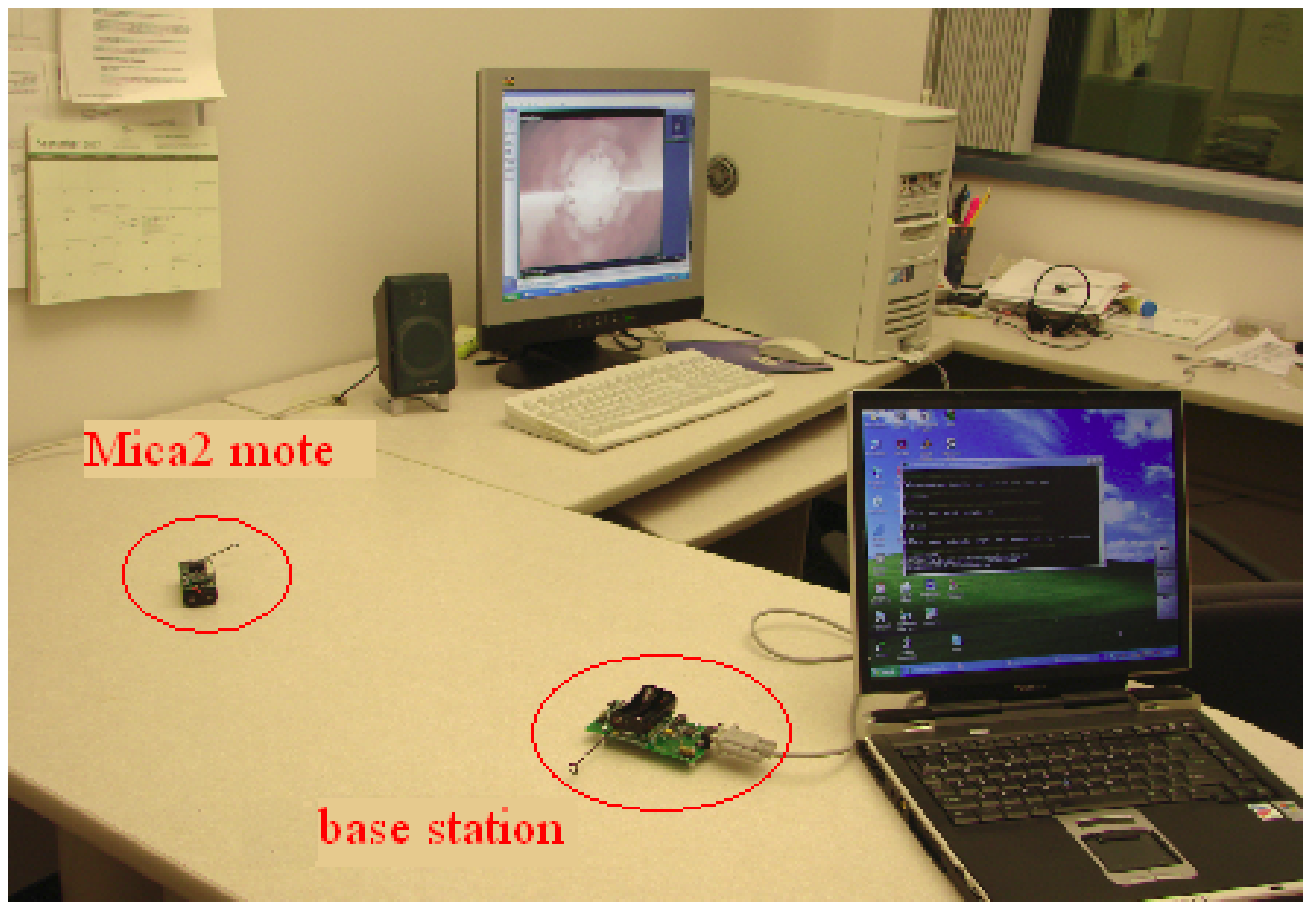}}
  \caption{The test bed using real sensors.} \label{fig:testbed}
\end{figure}

We implemented our online algorithms using MICA2\linebreak
motes~\cite{MICA2} from Crossbow Technology Inc. The test bed is
illustrated in Figure~\ref{fig:testbed}. A MICA2 mote includes a radio/processor board and a sensor board.
The radio/processor board uses $900$ Mhz radio. The sensor board
includes a microphone that can be used for sampling sound. The
interface of the base station is based on RS232. It acts as a gateway to connect
the laptop and the radio wireless sensor network. \nop{Our laptop
does not have an RS232 port and thus we use a USB/RS232 adaptor to
connect to the base station interface.} The original audio files are
played on a desktop computer, monitored by a MICA2 mote, and
transmitted over wireless channel from the MICA2 mote to the base
station. \nop{For details of program architecture, please refer to our
development documentation~\cite{Emad}.}

\looseness -1
The results about the sample reduction ratio on the real sensor test
bed are close to the simulation results using Matlab. But the audio
quality obtained using the real test bed is worse than that obtained
in the Matlab simulation. The deterioration in audio quality is
caused by the major restriction of TinyOS~\cite{Cull}, the current
operating system in MICA2 motes. The OS does not support multiple
threads and thus it cannot perform radio transmission and sound
sampling concurrently. Due to this limit, when we transmit data to
the base station, the sensor board stops sampling and the sound
during this period is missed, resulting in small silent gaps in the
recovered audio. Nevertheless, we can still recognize the human
speech and the piano music.

The same task can be carried out with the most recent, more
advanced sensor device, MICAz from the same company. With a higher
price, MICAz sensors support up to $250$ Kbps wireless transmission.
This task, however, has never been fulfilled with low-end devices
like MICA2. To this end, we break the limit of scarce radio
bandwidth and carry out a task that is hard to achieve without our
fast online compression methods.

\subsection{Evaluation in Other Applications}

Although we only implemented the online algorithms in an acoustic
sensor monitoring system, our algorithms are actually applicable to
many other application domains such as electrocardiogram (ECG)
monitoring for patients. We test our algorithm on an ECG data set
The maximum value on the data set is $2,490$ and the minimum value
is $-8,190$. We test our online algorithms with error bound varying
from $1$ to $100$.

\begin{figure}[htb!]
  \centerline{\includegraphics[width=60mm]{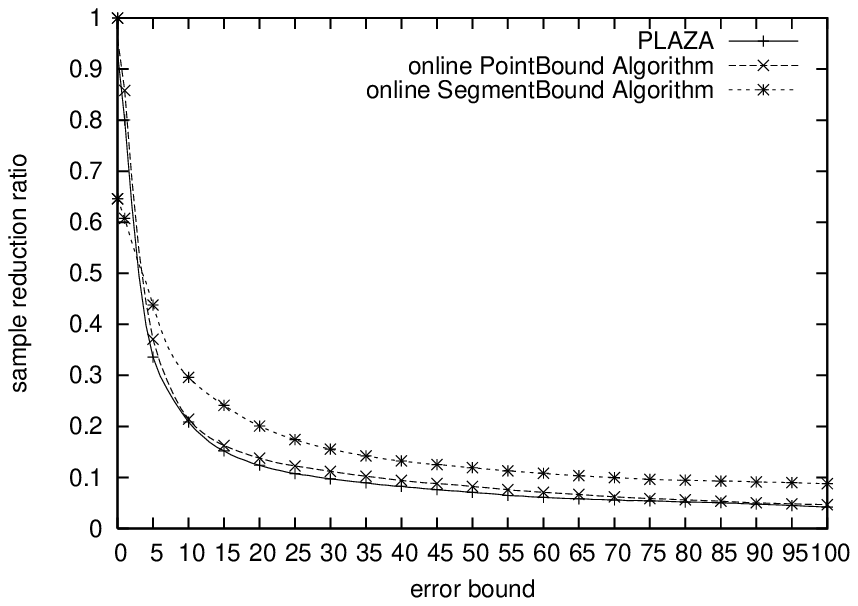}}
  \caption{Results on an ECG data set.} \label{fig:ECG}
\end{figure}

Figure~\ref{fig:ECG} compares the sample reduction ratio of
algorithms PLAZA, PointBound, and SegmentBound on the ECG data set.
The performance of algorithms PLAZA and PointBound is very similar.
When the error bound is set to over $35$, both algorithms can
compress the data up to $10\%$ of the original size. The gap between
algorithms SegmentBound and PointBound comes from the fact that,
using the same error bound, the PLA-SegmentBound problem and the
PLA-PointBound problem put a tighter error constraint on the
PLA-SegmentBound problem.

\section{Conclusion}
\label{sec:con}

In this paper, we tackle the problem of online compression of data
streams in the resource-constrained network environment, where the
traditional data compression techniques cannot apply. Particularly,
we aim at fast piecewise linear approximation (PLA) methods with
quality guarantee. We study two versions of the problem which
explore quality guarantees in different forms. For the error bounded
PLA problem, we design fast online algorithms running in linear time
complexity and requiring a constant space cost. The online
algorithms are also optimal in terms of the number of generated
segments. To meet the needs from tiny, resource-constrained sensors,
we develop another online algorithm that involves very simple
computation and generates connecting line segments. Our simulation
results and the test using a real sensor test bed demonstrate that
our fast online linear approximation methods are very effective for
data stream compression and transmission over low bandwidth networks
with nodes heavily constrained in computational power.

Equipped with the insights gained in this study, we see a lot
of application opportunities for our methods. Meanwhile, there are
also some interesting open questions for future work. For example,
an interesting question is to design an online algorithm that can
compute an $\epsilon$-PLA consisting of connecting line segments
that has an approximation factor to the optimum.
\section*{Acknowledgement}
This research was supported by Natural Sciences and Engineering Research Council of Canada (NSERC), Canada Foundation for Innovation (CFI), and the British Columbia Knowledge Development Fund (BCKDF). We also thank Dr.~E.~Keogh for his informative comments and for providing the ECG dataset.
\balance

\bibliographystyle{plain}
\small \baselineskip 9pt
\bibliography{streamcompression}

\end{document}